\documentclass[10pt,conference,letterpaper]{IEEEtran}
\usepackage{cite}
\usepackage{amsmath,amssymb,amsfonts}
\usepackage{algorithm}
\usepackage{algorithmic}
\usepackage{graphicx}
\usepackage{textcomp}
\usepackage{xcolor}
\usepackage[hyphens]{url}
\usepackage{enumitem}
\usepackage{balance}
\usepackage{subfigure}
\usepackage{multirow}
\usepackage{tabularx}
\usepackage{threeparttable}
\usepackage{float}
\usepackage{stfloats}
\usepackage{placeins}
\usepackage{booktabs}
\usepackage{array}
\usepackage[hypertexnames=false,hidelinks]{hyperref}

\graphicspath{{architecture-figures/}{data-figures/}}
\DeclareGraphicsExtensions{.pdf,.png,.jpg,.jpeg}
\newcommand{\archincludegraphics}[2][]{%
  \includegraphics[trim=1.5pt 1.5pt 1.5pt 1.5pt,clip,#1]{#2}}
\newcommand{\garblingfreepanel}[2]{%
  \begin{minipage}[c][0.22\columnwidth][c]{#1}
    \centering
    \archincludegraphics[scale=0.425,trim=1.5pt 1.5pt 1.5pt 12pt]{#2}
  \end{minipage}}
\newcolumntype{C}[1]{>{\centering\arraybackslash}p{#1}}
\newcolumntype{M}[1]{>{\centering\arraybackslash}m{#1}}
\DeclareRobustCommand{\rvinst}[1]{\texttt{#1}}
\DeclareRobustCommand{\cinterface}[1]{\texttt{#1}}
\DeclareRobustCommand{\statefield}[1]{\texttt{#1}}
\DeclareRobustCommand{\compvalue}[1]{\texttt{#1}}
\title{RV2GC: Reconstructing the Garbling Boundary\\
in a RISC-V Garbled Processor}
\author{
  \IEEEauthorblockN{
    Zixiao Feng\textsuperscript{1,2,3},
    Xiaolin Li\textsuperscript{1,2,3},
    Chenglu Jin\textsuperscript{4},
    Wei Yan\textsuperscript{1,2,3},\\
    Yong Zhang\textsuperscript{3},
    Hongwei Liu\textsuperscript{1,2},
    Yong Liu\textsuperscript{3},
    Qinfen Hao\textsuperscript{1,2,3}
  }
  \IEEEauthorblockA{
    \textsuperscript{1}Institute of Computing Technology, Chinese Academy of Sciences\\
    \textsuperscript{2}University of Chinese Academy of Sciences\\
    \textsuperscript{3}Zhongguancun Laboratory\\
    \textsuperscript{4}Centrum Wiskunde \& Informatica\\
    \texttt{\{fengzixiao24z,lixiaolin19s,yanwei,liuhongwei,haoqinfen\}@ict.ac.cn}\\
    \texttt{chenglu.jin@cwi.nl; \{zhangyong,liuyong\}@zgclab.edu.cn}
  }
}

\hypersetup{
  pdftitle={RV2GC: Reconstructing the Garbling Boundary in a RISC-V Garbled Processor},
  pdfauthor={Zixiao Feng, Xiaolin Li, Chenglu Jin, Wei Yan, Yong Zhang, Hongwei Liu, Yong Liu, Qinfen Hao}
}

\begin{document}
\maketitle
\thispagestyle{plain}
\pagestyle{plain}
\flushbottom
\setlength{\textfloatsep}{10pt minus 2pt}
\setlength{\floatsep}{8pt minus 2pt}
\setlength{\intextsep}{10pt plus 2pt minus 2pt}
\setlength{\dbltextfloatsep}{10pt minus 2pt}
\setlength{\dblfloatsep}{10pt minus 2pt}

\setlength{\abovecaptionskip}{2pt}
\setlength{\subfigbottomskip}{2pt}


\begin{abstract}
Programmable garbled-circuit execution must balance the benefits of cross-program reuse against the cost of repeated work during secure execution. A complete garbled processor provides a fixed circuit for many programs, but its evaluator repeatedly visits processor-wide structures even when only a small part performs the current protected computation. We present RV2GC, which reconstructs the garbling boundary around a fixed library of RISC-V instruction circuits. Public scheduling and operand routing operate outside the circuit topologies; persistent label-encoded state connects independently invoked garbled instruction circuits and instruction-specific garbling-free paths through a common interface. A Conditional-execution Stack and protected commits support secret-dependent branches, bounded loops, and secret-indexed memory access under public scheduling, keeping instruction and memory-address sequences independent of private inputs. Across nine benchmarks, measured ARM2GC-to-RV2GC runtime ratios range from 12.81 to 4,043. In terms of circuit size, the largest listed instruction circuit has 3,386 gates, corresponding to 2.044\% and 0.393\% of the 256-byte and 4,096-byte Amber processors, respectively. Additional comparisons quantify schedule storage, communication, and the cost of secret-dependent execution.
\end{abstract}


\section{Introduction}
\label{Introduction}

Garbled circuits (GCs) enable two parties to compute over private inputs while revealing the designated outputs~\cite{DBLP:conf/focs/Yao86}. Programmable execution must choose which structures remain reusable across programs and which are specialized. Complete garbled processors retain a fixed processor circuit across programs~\cite{DBLP:conf/sp/SonghoriHS0K15,DBLP:conf/dac/SonghoriRHSK19}. Function-specific circuits and program-specialized instruction structures instead encode more of the application into the secure execution artifact~\cite{DBLP:conf/cc/FranzHKSV14,DBLP:conf/eurosp/MoodGCBT16,DBLP:conf/esorics/WangGMK16,DBLP:conf/ccs/YangPHK23}. These choices affect preparation cost and repeated execution work.

ARM2GC illustrates the cost of a complete-processor boundary. Its SkipGate algorithm avoids cryptographic processing for gates that can be resolved from public information, but identifying those gates still visits the processor netlist~\cite{DBLP:conf/dac/SonghoriRHSK19}. Our profile of nine benchmarks shows that evaluator work other than protected-gate processing accounts for 71.3--93.2\% of ARM2GC's measured runtime, including netlist traversal and associated management. The observation motivates a structural question: can a programmable GC system keep a fixed cross-program circuit substrate while invoking only the computation needed for the current operation?

RV2GC answers this question by reconstructing the garbling boundary. Public instruction scheduling and operand routing are separated from the instruction computations that require garbled evaluation. Register and memory state are stored outside the individual circuit topologies, with secret values retained as wire labels. A fixed library of RISC-V instruction circuits supplies the protected operations. Program changes within the supported ISA change the public schedule and initialization data, while the library topologies remain unchanged.

During execution, RV2GC connects instruction circuits and garbling-free paths through persistent label-encoded state to form a continuous execution. It targets oblivious programs whose public execution traces are independent of private inputs. The Conditional-execution Stack and protected commits allow secret-dependent choices to affect state updates under this public schedule.

Our two-party software implementation is evaluated against ARM2GC, Wang~\cite{DBLP:conf/esorics/WangGMK16}, VISA~\cite{DBLP:conf/ccs/YangPHK23}, and function-specific circuits. We examine runtime, communication, stateful execution, and program storage. We also use BINSEC~\cite{DanielBR23Binsec} to check that control flow and memory addresses in the final benchmark binaries are independent of private inputs; Section~\ref{sec:security-discussion} presents the methods and results.

We make three technical contributions:

\begin{itemize}
  \item \textbf{Garbling-boundary reconstruction.}
  We rethink which components of programmable execution should reside within the repeatedly traversed garbled-circuit topology. RV2GC separates public instruction scheduling and operand routing from garbled computation, decouples label-encoded architectural state from individual circuit topologies, and implements instruction operations requiring garbled evaluation through a fixed library of RISC-V instruction circuits. This reconstruction preserves cross-program reuse of circuit topologies while limiting online circuit traversal to the circuitry required by the scheduled operation.

  \item \textbf{Stateful instruction composition with garbling-free execution.}
  We design a persistent state interface that composes separately invoked instruction circuits and garbling-free execution paths into continuous RV32 execution. Public register and memory values are maintained explicitly, while protected values persist as wire labels under a common two-party label-state invariant. Both execution paths preserve this invariant, allowing intermediate results to pass between instructions without decoding protected values. For eligible instruction cases, public computation, label-domain manipulation, preparation-time resolution, or direct state movement realizes the primary instruction semantics without encrypted garbled tables.

  \item \textbf{Secret-dependent execution under public scheduling.}
We design a Conditional-execution Stack and protected commits that support secret-dependent branches, bounded loops, and secret-indexed memory access under a public instruction and address schedule. The stack maintains effective protected predicates across nested conditional regions; commits use these predicates to select candidate state updates or preserve prior values. This interface carries secret-dependent choices through protected state while the public schedule remains independent of private inputs.
\end{itemize}

\section{Background and Motivation}
\label{idea}

\subsection{Garbled Circuits and Programmable Boundaries}
\label{sec:gcprotocol}
\label{sec:gcopt}

Yao's garbled-circuit (GC) protocol enables two parties to evaluate a Boolean circuit over private inputs while revealing only the designated output~\cite{DBLP:conf/focs/Yao86}. The generator assigns two opaque labels to each wire, encodes gate functionality in garbled tables, and supplies labels for its own inputs; the evaluator obtains its input labels through oblivious transfer (OT), propagates the active labels gate by gate, and decodes only authorized outputs. Free-XOR, fixed-key garbling, and half-gates reduce gate-level cost~\cite{DBLP:conf/icalp/KolesnikovS08,DBLP:conf/ccs/GueronLNP15,DBLP:conf/eurocrypt/ZahurRE15}, but do not determine which execution structures lie inside a reusable garbling boundary. Sequential GC carries label-encoded state across cycles, and TinyGarble uses this model to realize complete garbled processors~\cite{DBLP:conf/sp/SonghoriHS0K15}.

Table~\ref{tab:boundary-design-space} compares five boundary organizations across per-program artifacts, fixed structures, and online paths. CBMC-GC and Frigate use function-specific circuits~\cite{DBLP:conf/cc/FranzHKSV14,DBLP:conf/eurosp/MoodGCBT16}; Wang and VISA derive program-specific instruction or fragment structures~\cite{DBLP:conf/esorics/WangGMK16,DBLP:conf/ccs/YangPHK23}; ARM2GC retains a complete processor circuit traversed with SkipGate~\cite{DBLP:conf/dac/SonghoriRHSK19}; and RV2GC retains an instruction-circuit library and persistent-state interface, with a public schedule selecting operations.

\begin{table}[!tb]
  \footnotesize
  \centering
  \setlength{\tabcolsep}{1.5pt}
  \renewcommand{\arraystretch}{1.08}
  \caption{Comparison of programmable GC systems by per-program artifacts, fixed structures, and online execution paths.}
  \label{tab:boundary-design-space}
  \renewcommand{\tabularxcolumn}[1]{m{#1}}
  \begin{tabularx}{\columnwidth}{@{}>{\raggedright\arraybackslash}m{0.16\columnwidth}*{3}{>{\raggedright\arraybackslash}X}@{}}
    \toprule
    \textbf{System} & \textbf{Per-Program} & \textbf{Fixed Structure} & \textbf{Online Execution} \\
    \midrule
    CBMC-GC / Frigate &
    Function-specific circuit &
    GC compiler; GC backend &
    Function-circuit evaluation \\
    \midrule
    Wang &
    Instruction banks; specialized CPU circuits &
    MIPS state interface; ORAM primitives &
    Oblivious fetch; CPU/ORAM execution \\
    \midrule
    VISA &
    Fragment circuits; active-set stacks &
    Stacked Garbling; EpiGRAM &
    Fragment selection/evaluation; EpiGRAM access \\
    \midrule
    ARM2GC &
    Program binary &
    Complete processor circuit &
    Processor-wide traversal \\
    \midrule
    RV2GC &
    Public instruction schedule &
    Instruction-circuit library &
    Scheduled-operation execution \\
    \bottomrule
  \end{tabularx}
\end{table}

The table separates what remains fixed across programs from what is traversed at each step. RV2GC uses a program-specific public schedule to select and connect operations from a fixed library.

\subsection{Cost of a Complete-Processor Boundary}
\label{Performance}

SkipGate classifies gates using public values and wire relationships, then skips unnecessary garbling or evaluation~\cite{DBLP:conf/dac/SonghoriRHSK19}. In the ARM2GC implementation we measure, this classification visits the processor netlist each cycle. Fetch, decode, register selection, and other processor structures therefore contribute evaluator work even when their values are publicly determined.

Our profile attributes 1.0--4.2\% of measured runtime to protected-gate processing and 71.3--93.2\% to the remaining evaluator work, including traversal, public-value propagation, state copying, and gate management. RV2GC reduces repeated access to the complete processor structure by directly dispatching the instruction computation needed for the current operation.

\subsection{Execution and Threat Model}
\label{sec:threat-model}

RV2GC uses a public-function, semi-honest two-party model: Alice garbles and Bob evaluates, and both follow the protocol. We assume secure GC and oblivious transfer (OT).

Both parties share the program, fixed instruction-circuit topologies, and public configuration: public inputs and execution-size parameters fixed independently of private inputs. These determine the public schedule, operand routing, and memory-address sequence.

For a supported RV32 binary $B$, public configuration $u$, and private input $s$, let $\operatorname{Tr}(B,u,s)$ denote its logical instruction-control and memory-address trace. We call $B$ an \emph{oblivious program} under $u$ when, for all allowed private inputs $s_1,s_2$,
\begin{equation}
\operatorname{Tr}(B,u,s_1)=\operatorname{Tr}(B,u,s_2).
\label{eq:trace-contract}
\end{equation}
The computed results and logical state updates may still depend on private inputs.

The parties' private inputs and protected intermediate state are the objects of protection; designated outputs are revealed only to authorized recipients. The privacy goal is that protocol execution reveals no additional private information beyond what a party can learn from the public information, its own input, and its authorized outputs.

\section{RV2GC Architecture and Implementation}
\label{sec:design}

\subsection{System Overview}
\label{RV2GC Framework}
\label{sec:system-overview}

\begin{figure}[!tb]
  \centering
  \archincludegraphics[width=0.95\columnwidth,trim=1.5pt 15pt 1.5pt 15pt]{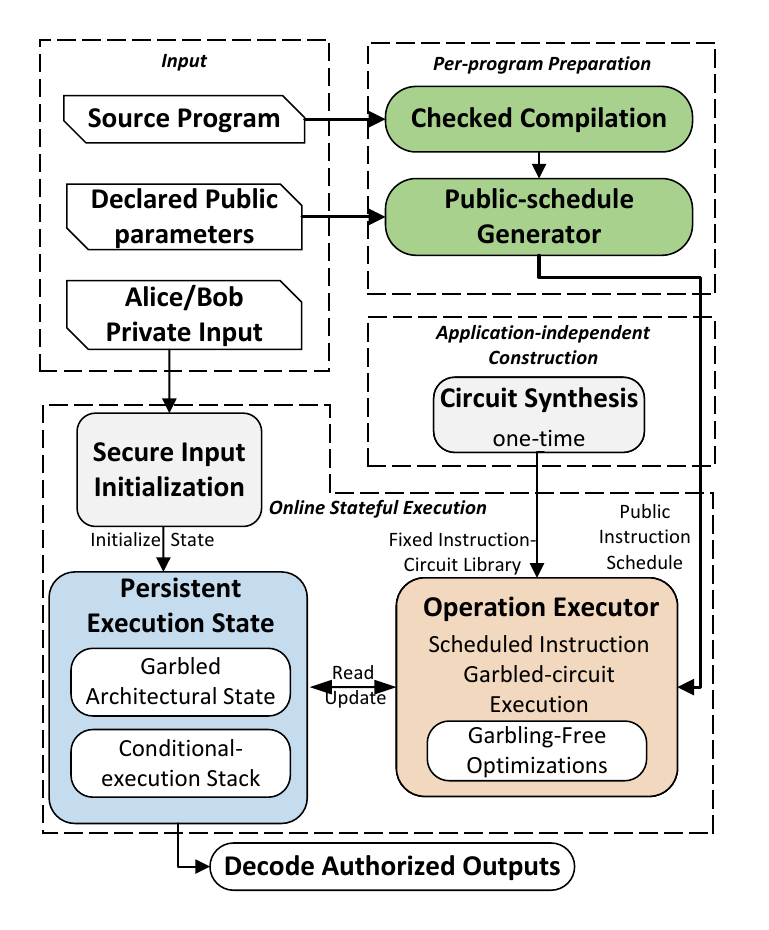}
\caption{RV2GC's per-program preparation, application-independent circuit construction, and online stateful execution. Prepared programs share the public-schedule generator, instruction-circuit library, and runtime.}
  \label{fig:system-overview}
\end{figure}

RV2GC runs as software on general-purpose CPUs. Alice and Bob are separate processes connected over TCP. HDL synthesis produces Boolean instruction netlists for software garbling and evaluation; it does not produce a hardware accelerator. Figure~\ref{fig:system-overview} shows per-program preparation, application-independent circuit construction, and online stateful execution.

The library is synthesized once. Per-program preparation produces an oblivious RV32 binary, its public instruction schedule, and initialization data. The prepared programs use the same runtime and circuit library. Section~\ref{sec:program-preparation} describes the preparation paths and their checks.

At execution time, input encoding and OT initialize public values and secret labels in registers and memory. The Operation Executor follows the schedule, reads the Persistent Execution State, and invokes either an instruction circuit or a garbling-free path. The Conditional-execution Stack carries protected condition state across these operations. Only designated outputs are decoded.

\subsection{Reconstructing the Garbling Boundary}
\label{sec:library-dispatch}

\begin{figure}[!tb]
  \centering
  \archincludegraphics[width=\columnwidth,trim=1.5pt 20pt 1.5pt 22pt]{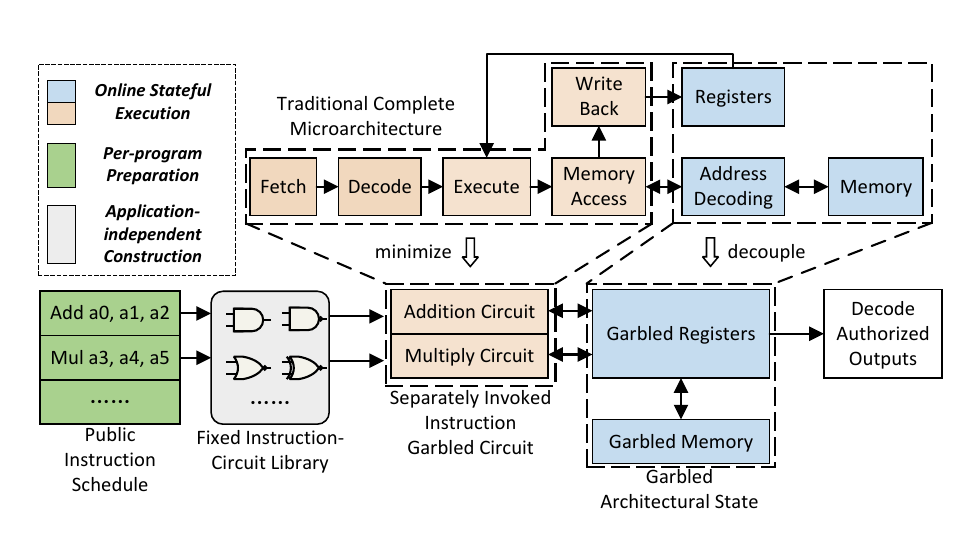}
  \vspace{-2pt}
  \caption{Reconstructing the garbling boundary. A complete garbled microarchitecture encloses processor-wide control, execution, and state structures. RV2GC instead selects from a fixed instruction-circuit library and decouples the Garbled Architectural State from the complete processor.}
  \label{fig:boundary-comparison}
\end{figure}

Figure~\ref{fig:boundary-comparison} contrasts two garbling-boundary organizations. A complete garbled microarchitecture encloses fetch, decode, execution, writeback, registers, address decoding, and memory in one processor-wide topology. RV2GC makes two architectural changes: it \emph{minimizes} the protected structure invoked for an operation to the circuitry needed to implement its instruction semantics, and it \emph{decouples} register and memory state from the complete processor circuit. The result is a library of separately invoked instruction circuits connected through Persistent Execution State.

The fixed library contains the protected instruction circuits for the supported RV32 subset, synthesized once from HDL into TinyGarble-compatible SCD format. Each public schedule entry identifies an operation and the routing information that connects its operands and result to architectural state; protected operands remain in that state rather than in the schedule.

The schedule is derived from an RV32 binary satisfying Eq.~\ref{eq:trace-contract} and its declared public parameters. It remains unchanged when only private inputs change; a different program or public configuration may require a new schedule. Schedule generation must preserve the binary trace when producing operation and routing metadata.

At each online step, both parties consume the same schedule entry and select the same implementation from the fixed library. When an operation requires non-free gates, its current state labels become circuit inputs and invocation-specific labels and garbled tables are produced for that circuit. The output labels return to persistent state for later operations. For eligible operations, instruction-specific garbling-free optimizations realize their semantics without encrypted garbled tables, as described below. RV2GC therefore reuses circuit \emph{topologies}---not pre-garbled instances---and restricts protected work to the scheduled operation rather than a complete processor traversal.

\subsection{Stateful Instruction Composition with Garbling-Free Execution}
\label{sec:stateful-instruction-execution}

Separately invoked instruction circuits must preserve a consistent representation of register and memory state.

\subsubsection{Persistent Execution State Across Operations}
\label{sec:persistent-label-state}

Separately invoked operations advance RV2GC's Persistent Execution State, which has two parts. The Garbled Architectural State comprises Garbled Registers and Garbled Memory in persistent software-managed arrays outside the complete processor circuit; each bit is either an explicit public value or a protected wire-label representation. The Conditional-execution Stack retains conditional context across scheduled operations and is detailed in Section~\ref{sec:secret-dependent-execution}. Figure~\ref{fig:boundary-comparison} depicts only the protected portion of the Garbled Architectural State. Public schedule metadata selects register and memory locations without decoding protected values. Only values classified as public enter public computation; protected operands retain their label representation along the routes specified by the schedule.

At schedule boundary $t$, every protected state bit $b$ with logical value $v_b$ satisfies a two-party label-state invariant: Alice stores a pair $(K^0_{t,b},K^1_{t,b})$, Bob stores only the active label $K^{v_b}_{t,b}$, and the session-wide Free-XOR offset $R$ gives $K^1_{t,b}=K^0_{t,b}\oplus R$. Input encoding and OT initialize register and memory representations; stack transitions derive valid labels for conditional state.

A garbled instruction circuit returns a new pair/active-label correspondence; copies and bit-position permutations apply a common public mapping; XOR and NOT preserve the offset through label algebra; and public operations produce explicit public results. Conditional-execution Stack transitions and protected selection preserve the same invariant. Induction over the public schedule therefore shows that each operation preserves the label-state relation, enabling continuous execution without decoding intermediate secrets. Persistent memory arrays also decouple logical memory capacity from the fixed topology: larger memories require additional state storage, not a larger synthesized processor circuit.

\subsubsection{Garbling-Free Instruction Execution}

\begin{figure}[!tb]
  \centering
  \subfigure[XOR/NOT]{\label{fig:garbling-free-xor-not}\archincludegraphics[width=0.90\columnwidth]{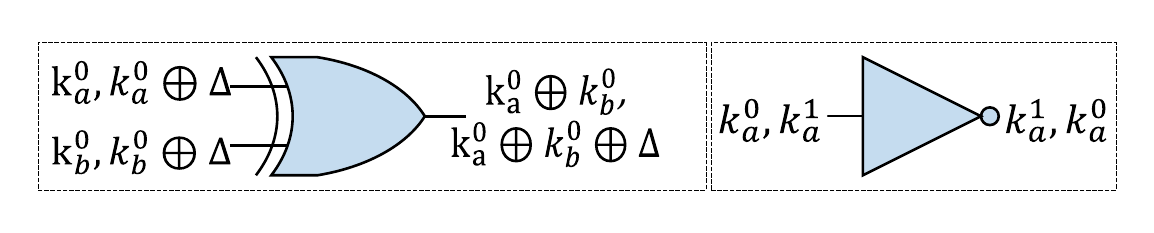}}\\
  \subfigure[Shift]{\label{fig:garbling-free-shift}\garblingfreepanel{0.26\columnwidth}{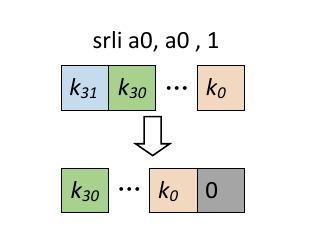}}\hspace{0.02\columnwidth}
  \subfigure[Long Immediate]{\label{fig:garbling-free-long-immediate}\garblingfreepanel{0.48\columnwidth}{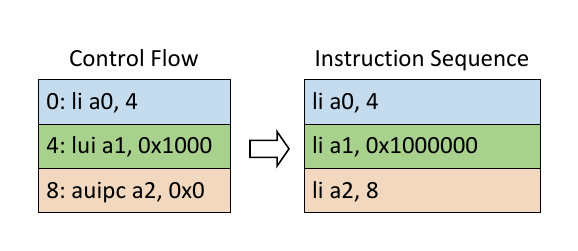}}\\
  \subfigure[Branch/Jump]{\label{fig:garbling-free-branch}\garblingfreepanel{0.48\columnwidth}{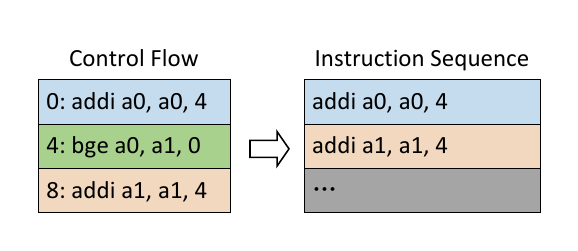}}\hfill
  \subfigure[Memory Access]{\label{fig:garbling-free-memory}\garblingfreepanel{0.48\columnwidth}{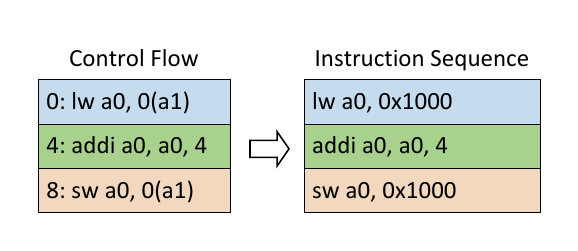}}
  \caption{Five garbling-free instruction types. (a) XOR and NOT operate directly in the label domain; (b) a fixed-amount logical shift permutes labels; (c) long-immediate values are resolved during preparation; (d) public branch and jump outcomes determine the public schedule; and (e) public memory addresses enable direct state transfer.}
  \label{fig:garbling-free-paths}
\end{figure}

Within the operation-level garbling boundary, some instruction cases admit garbling-free implementations. RV2GC applies garbling-free optimizations when public computation, label-domain manipulation, preparation-time resolution, or direct state movement can realize the instruction semantics while preserving the Garbled Architectural State representation used by scheduled instruction execution. We call this handling \emph{garbling-free} because the instruction's primary semantics require no encrypted garbled-table entries to be generated, transmitted, or evaluated. When all operands are explicit public values, both parties can evaluate the operation locally. This case applies across multiple instruction classes. Figure~\ref{fig:garbling-free-paths} shows five instruction classes to which RV2GC applies such optimizations.

\noindent\textbf{XOR and NOT.}
RV2GC applies the standard Free-XOR construction~\cite{DBLP:conf/icalp/KolesnikovS08} to XOR. Let protected input wires $x$ and $y$ carry bits $a$ and $b$, with each wire $w$ satisfying $K_w^1=K_w^0\oplus R$. Alice derives $K_z^0=K_x^0\oplus K_y^0$ and $K_z^1=K_z^0\oplus R$, while Bob derives the active output label by computing $K_x^a\oplus K_y^b$. Because
\[
K_x^a\oplus K_y^b
=K_x^0\oplus K_y^0\oplus(a\oplus b)R
=K_z^{a\oplus b}
\]
the resulting label pair and active label represent $a\oplus b$ under the same label-state invariant. For NOT, Alice exchanges the semantic roles of the two labels while Bob retains the same active label under the exchanged interpretation, producing a valid output representation without an encrypted garbled table.

\noindent\textbf{Fixed-amount logical shift.}
A logical shift by a public immediate applies the same public bit-position permutation to Alice's label pairs and Bob's active labels. Vacated positions become explicit public zeros. This implements the shift without invoking a circuit.

\noindent\textbf{Long-immediate instructions.}
The results of \texttt{lui}, \texttt{auipc}, and the \texttt{li} pseudo-instruction are determined by the binary layout and public execution point. Per-program preparation resolves these values and records explicit public immediate-load operations in the schedule, so no online circuit is needed to reconstruct them.

\noindent\textbf{Public control-flow instructions.}
When a branch or jump outcome is public, the public-schedule generator follows the selected target and records the resulting instruction sequence. The control-transfer instruction itself need not be dispatched during online execution.

\noindent\textbf{Public-address memory instructions.}
When an effective memory address is public, the schedule identifies the location and the runtime transfers representations directly between persistent register and memory arrays. Loads copy the selected memory representation to a register, while stores copy the source representation to memory. Narrow signed loads replicate the sign label and unsigned loads insert public zeros in the upper positions. No in-circuit address decoder or encrypted table is required.

\subsection{Secret-Dependent Execution under Public Scheduling}
\label{sec:secret-dependent-execution}

Secret-dependent branches, loop termination, and array indices ordinarily affect the execution path or accessed addresses. RV2GC expresses these choices through protected values while a public schedule fixes the instruction and address sequences. The schedule determines which computations occur; protected predicates determine which results take effect in program state.

\subsubsection{Conditional-execution Stack and Protected Commits}

The Conditional-execution Stack carries the effective predicate of the current conditional region across instruction invocations. Each scheduled operation computes a candidate result according to its RV32 semantics. In a conditional region, a protected MUX uses the predicate to select the candidate or the prior destination value; otherwise, the candidate directly updates its destination. All scheduled computations must remain defined and their memory accesses valid, including when their updates are suppressed.

Three instructions in the RV32 custom-instruction space---\rvinst{pushbr}, \rvinst{popbr}, and \rvinst{updatebr}---maintain this conditional context. They update a Conditional-execution Stack that persists across scheduled operations. Each stack level contains the following three state fields:

\begin{itemize}[leftmargin=1.35em,labelsep=0.45em,itemsep=1pt,topsep=2pt,parsep=0pt]
  \item \statefield{BRANCH\_STATE}: A public mode bit that indicates whether the current operation is under conditional execution.
  \item \statefield{BRANCH\_COND}: A label-encoded effective predicate that determines whether updates in the current conditional region take effect, incorporating enclosing predicates when nested.
  \item \statefield{BRANCH\_ELSE}: A label-encoded predicate that records the remaining else condition used by a subsequent \texttt{else-if} or \texttt{else} branch.
\end{itemize}

The low-level C interface maps conditional constructs to the RV32 conditional-execution extension in Table~\ref{tab:conditional-execution-interface}. The mapped instructions update the Conditional-execution Stack, providing state transitions for protected secret-dependent execution.

\begin{table}[!tb]
\caption{C-interface mappings to RV32 conditional-execution instructions and Conditional-execution Stack operations.}
\label{tab:conditional-execution-interface}
\footnotesize
\centering
\setlength{\tabcolsep}{2pt}
\renewcommand{\arraystretch}{1.08}
\begin{tabularx}{\columnwidth}{@{}M{0.22\columnwidth}M{0.18\columnwidth}>{\raggedright\arraybackslash}X@{}}
  \toprule
  \textbf{C Interface} & \textbf{RV32 Extension} & \textbf{Stack Operation} \\
  \midrule
  \cinterface{RVIF(c)} & \rvinst{pushbr} & Push a new level: set \statefield{BRANCH\_STATE}=1; derive \statefield{BRANCH\_COND} and \statefield{BRANCH\_ELSE} from $c$ and $\neg c$, respectively; combine both with the enclosing \statefield{BRANCH\_COND} when nested. \\
  \midrule
  \cinterface{RVENDIF} & \rvinst{popbr} & Pop the top level and restore the enclosing branch context. \\
  \midrule
  \cinterface{RVELSEIF(c)} & \rvinst{updatebr} & Replace the top \statefield{BRANCH\_COND} and \statefield{BRANCH\_ELSE} using the remaining else condition combined with $c$ and $\neg c$, respectively. \\
  \midrule
  \cinterface{RVELSE} & \rvinst{updatebr} & Promote the top \statefield{BRANCH\_ELSE} to the current \statefield{BRANCH\_COND}. \\
  \bottomrule
\end{tabularx}
\end{table}

\begin{figure}[!tb]
  \centering
  \archincludegraphics[width=\columnwidth,trim=1.5pt 20pt 1.5pt 15pt]{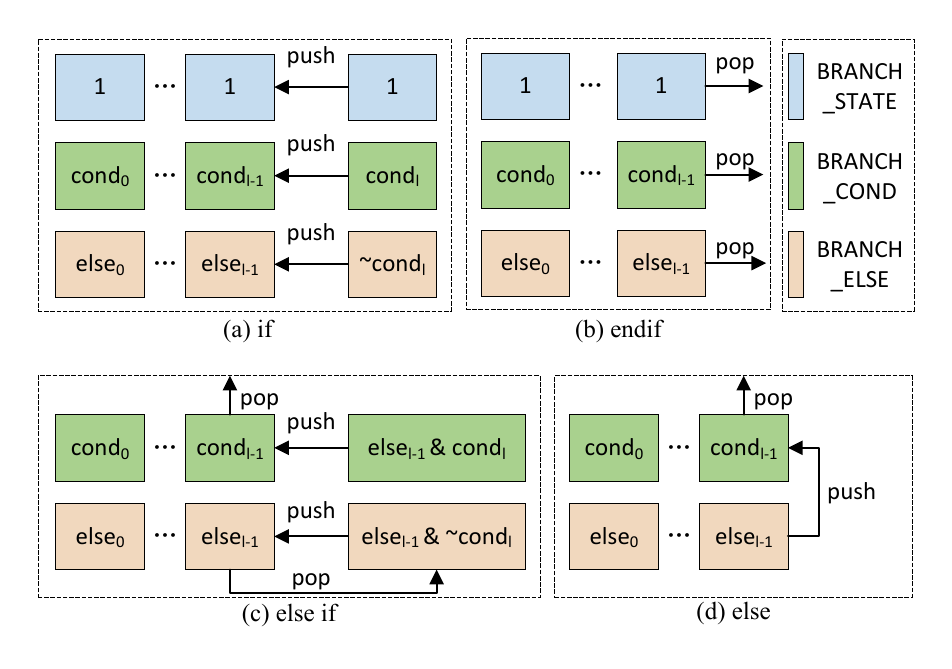}
  \vspace{-2pt}
  \caption{Conditional-execution Stack transitions for (a) entering an \texttt{if} region, (b) leaving the region, (c) advancing to \texttt{else-if}, and (d) advancing to \texttt{else}.}
  \label{fig:branch-state-updates}
\end{figure}

Figure~\ref{fig:branch-state-updates} visualizes these transitions. At any nesting depth, the current top \statefield{BRANCH\_COND} represents the effective protected predicate for that conditional region. \rvinst{pushbr} creates a new level and combines its condition with the enclosing predicate, \rvinst{updatebr} advances that level to an \texttt{else-if} or \texttt{else} case, and \rvinst{popbr} restores the enclosing context.

The same predicate-and-commit interface supports the three execution patterns below. Section~\ref{sec:program-preparation} describes how the program transformation tool generates these forms.

\subsubsection{Secret-Dependent Branches}

For a secret-dependent branch, the public schedule includes the operations from each branch body. Figure~\ref{fig:protected-branch} shows how protected commits make only the selected body's updates take effect. The stack combines each branch condition with its enclosing predicate, so nested regions obey the same rule without changing the public sequence.

\subsubsection{Publicly Bounded Secret Loops}

When loop termination depends on a secret, stopping execution directly on that condition would make the public instruction-sequence length reflect the secret-dependent iteration count. RV2GC therefore uses a public loop bound for the declared public configuration and requires it to cover the supported executions. The public schedule specifies that fixed number of iterations. Protected liveness records whether the source loop has logically terminated. Once it has, later iterations remain publicly scheduled but no longer commit the loop body's state updates.

\begin{algorithm}[!b]
\caption{Secret-dependent loop execution with a public bound.}
\label{alg:publicly-bounded-secret-loop}
\normalsize
\begin{algorithmic}[1]
  \raggedright
  \STATE $\compvalue{LOOP\_LIVE}_0 \leftarrow$ current execution condition
  \FOR{$i$ from 0 to public loop bound $-1$}
    \STATE $\mathit{Cond}_i \leftarrow$ evaluate the loop condition\\
      \hspace*{1.5em}using protected values
    \STATE $\compvalue{LOOP\_ACTIVE}_i \leftarrow \compvalue{LOOP\_LIVE}_i \land \mathit{Cond}_i$
    \STATE $\statefield{BRANCH\_COND} \leftarrow \compvalue{LOOP\_ACTIVE}_i$
    \STATE execute the fixed public body schedule
    \STATE $\compvalue{LOOP\_LIVE}_{i+1} \leftarrow \compvalue{LOOP\_ACTIVE}_i$
  \ENDFOR
\end{algorithmic}
\end{algorithm}

\begin{figure}[!tb]
  \centering
  \subfigure[Secret-Dependent Branch]{\label{fig:protected-branch}\archincludegraphics[width=0.96\columnwidth,trim=15pt 17pt 18pt 22pt]{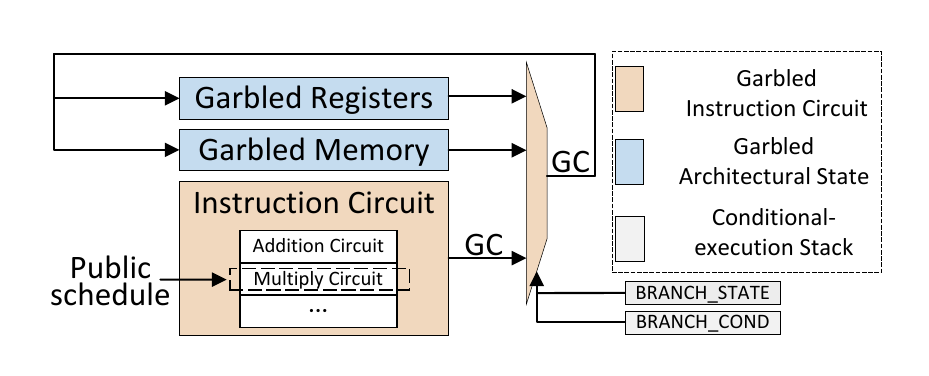}}
  \subfigure[Secret-Dependent Loop]{\label{fig:protected-loop}\archincludegraphics[width=0.96\columnwidth,trim=15pt 20pt 18pt 19pt]{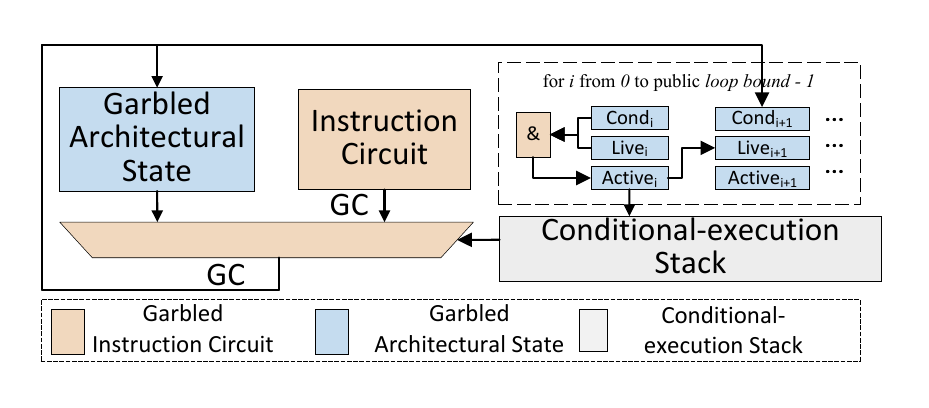}}
  \subfigure[Secret-Indexed Memory Access]{\label{fig:protected-memory}\archincludegraphics[width=0.96\columnwidth,trim=15pt 21pt 18pt 26pt]{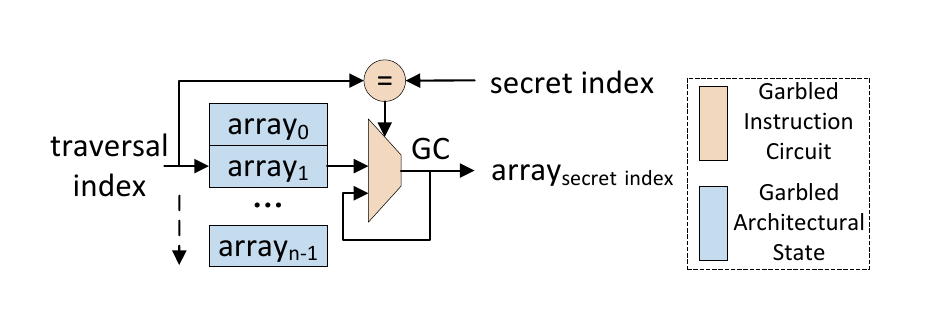}}
  \caption{Three protected execution patterns under public scheduling. (a) \statefield{BRANCH\_STATE} selects direct or conditional update; on the conditional path, \statefield{BRANCH\_COND} selects the candidate result or the current destination value. (b) A publicly bounded loop uses protected condition and liveness to control state updates. $\mathit{Cond}_i$, $\mathit{Live}_i$, and $\mathit{Active}_i$ correspond to the protected source condition, $\compvalue{LOOP\_LIVE}_i$, and $\compvalue{LOOP\_ACTIVE}_i$, respectively. (c) A fixed public-address scan uses protected equality and MUX feedback to select the secret-indexed entry.}
  \label{fig:protected-execution-patterns}
\end{figure}

Figure~\ref{fig:protected-loop} shows the state flow, and Algorithm~\ref{alg:publicly-bounded-secret-loop} gives the execution procedure. At iteration $i$, $\compvalue{LOOP\_LIVE}_i$ records whether the source loop remains active, and $\mathit{Cond}_i$ is its protected condition. Their conjunction, $\compvalue{LOOP\_ACTIVE}_i = \compvalue{LOOP\_LIVE}_i \land \mathit{Cond}_i$, governs the current updates. Carrying this active predicate into the next iteration keeps a logically terminated loop inactive.

\subsubsection{Secret-Indexed Access within a Public Block}

Directly addressing memory with a secret index would make the address sequence depend on that index. RV2GC instead visits every entry of a public address range that covers the allowed indices, denoted as $\mathit{array}_0,\ldots,\mathit{array}_{n-1}$. The range may be an array subrange; its addresses are public, while its entries and the selected index may be protected.

Figure~\ref{fig:protected-memory} shows the load path, and Algorithm~\ref{alg:secret-indexed-access} gives both load and store procedures. Each scan compares the public traversal index with the protected secret index. A load uses the equality result to select the matching value; a store updates the matching entry and preserves the others. Conditional stores combine the match with the current execution predicate.

\begin{algorithm}[!tb]
\caption{Secret-indexed loads and stores.}
\label{alg:secret-indexed-access}
\normalsize
\noindent\textbf{Secret-indexed load}
\begin{algorithmic}[1]
  \raggedright
  \STATE selected value $\leftarrow$ fixed public initializer
  \FOR{$i$ from 0 to $n-1$}
    \STATE match $\leftarrow$ protected equality($i$, secret index)
    \STATE selected value $\leftarrow$ protected MUX(match,\\
      \hspace*{1.5em}$\mathit{array}[i]$, selected value)
  \ENDFOR
\end{algorithmic}
\vspace{1pt}
\noindent\textbf{Secret-indexed store}
\begin{algorithmic}[1]
  \raggedright
  \FOR{$i$ from 0 to $n-1$}
    \STATE match $\leftarrow$ protected equality($i$, secret index)
    \STATE update condition $\leftarrow$\\
      \hspace*{1.5em}current execution condition $\land$ match
    \STATE $\mathit{array}[i] \leftarrow$ protected MUX(update condition,\\
      \hspace*{1.5em}store value, $\mathit{array}[i]$)
  \ENDFOR
\end{algorithmic}
\end{algorithm}

For a valid index, the value of the unique matching entry replaces the load's fixed public initializer and determines the result. The addresses traversed depend only on the configured range. Each access scans that range once, so its work grows with the number of entries; Section~\ref{sec:protected-scaling} evaluates this behavior.

\section{Correctness and Security}
\label{sec:security-discussion}

RV2GC must preserve the prepared program's result and protect private information under the model in Section~\ref{sec:threat-model}. We first explain how the execution mechanisms in Section~\ref{sec:design} meet these requirements, then describe the checks on program transformation and final binaries.

\subsection{Functional Correctness}

We take the prepared RV32 program as the reference execution. Input encoding establishes the correspondence between its initial logical values and their representation in Persistent Execution State. The instruction circuits and garbling-free paths in Section~\ref{sec:stateful-instruction-execution} implement the corresponding RV32 operations while preserving the representation required by subsequent operations.

The protected-update rules in Section~\ref{sec:secret-dependent-execution} make candidate updates take effect only under the effective predicate and preserve the prior destination otherwise. After a loop logically terminates, the remaining public iterations no longer commit the loop body's updates, preserving its result. For an index within the declared range, protected selection reads or updates the unique matching entry. Under the execution conditions stated with those mechanisms, these transitions maintain the logical-state correspondence throughout the scheduled execution; final decoding therefore yields the prepared program's specified outputs.

We further check the implementation of these mechanisms through complete-output comparisons. In the transformation tests, the original and transformed RV32 programs running on Spike produce the same complete outputs as an independent reference interpreter. Representative programs also yield matching decoded outputs in the Alice/Bob runtime. Section~\ref{sec:program-preparation} describes the test programs and comparison methods.

\subsection{Privacy Argument}

Equation~\ref{eq:trace-contract} requires identical logical instruction and address traces for different private inputs under a fixed program and public configuration. The schedule generator and runtime follow the public routing and value-classification rules in Section~\ref{sec:design}. Thus, public operation selection and memory access introduce no additional secret dependence into the observable execution sequence.

Under the GC/OT assumptions in Section~\ref{sec:threat-model}, input transfer and circuit evaluation protect the encoded private values. The invocation-specific garbling in Section~\ref{sec:library-dispatch} and the label-state interface in Section~\ref{sec:persistent-label-state} connect successive operations without decoding their protected predicates or intermediate values. Garbling-free operations use the public mappings and label algebra specified in that interface, and protected commits retain encoded results for later instructions. With these operations and routing rules correctly implemented, the system decodes only designated outputs. The resulting composition reveals no additional private information beyond a party's own input, the public information defined in Section~\ref{sec:threat-model}, and its authorized outputs.

\subsection{Program Transformation and Functional Validation}
\label{sec:program-preparation}

RV2GC supports directly written oblivious programs and provides a transformation tool that prepares supported C programs for the same RV32 runtime. Within a single C translation unit, the tool tracks dependencies on private inputs. It replaces secret-dependent branching with operations executed in a fixed public order, using protected conditions to make only the selected branch's state updates take effect. It converts secret-terminating loops to execution with public bounds and secret-indexed accesses to fixed scans over public ranges.

Preparation checks that public loop bounds and counter forms are identifiable and that scan objects and their extents are resolvable and consistent. The tool analyzes dependencies in GCC's intermediate representation (GIMPLE/SSA) to check for ordinary branches or memory addresses that still depend on secrets after transformation. Strict checking rejects unresolved pointer targets and external calls.

To preserve an explicit conditional-commit structure through compilation, the tool first computes a candidate value, then uses a protected condition to select it or retain the previous value. Each generated commit span enters conditional execution with \rvinst{pushbr}, performs one register copy (\rvinst{mv} or \rvinst{c.mv}), and exits with \rvinst{popbr}. Before and after linking, the validator checks that these entry and exit instructions are paired, that each span contains only the intended copy, and that the number of spans matches the transformation result. This constraint applies to automatically generated commit spans; directly written programs can still use the nested conditional execution described in Section~\ref{sec:secret-dependent-execution}. Preparation stops if transformation or checking fails.

To check that transformation preserves program results, we use the 34 test programs in Table~\ref{tab:transformation-tests}. They cover combinations of branches, loops, and secret-indexed accesses, as well as concrete computations such as conditional swaps and searches.

\begin{table}[!tb]
\centering
\footnotesize
\setlength{\tabcolsep}{3pt}
\renewcommand{\arraystretch}{1.1}
\caption{Test programs for validating program transformation.}
\label{tab:transformation-tests}
\renewcommand{\tabularxcolumn}[1]{m{#1}}
\begin{tabularx}{\columnwidth}{@{}>{\raggedright\arraybackslash}m{0.35\columnwidth}>{\raggedright\arraybackslash}X@{}}
\toprule
\textbf{Item} & \textbf{Contents} \\
\midrule
\multicolumn{2}{@{}l}{\emph{Mechanism combinations}} \\
\addlinespace[2pt]
Branches & None; single-level; nested \\
Loops & None; bounded \texttt{for}; bounded \texttt{while} \\
Secret-indexed memory & None; read; write \\
\midrule
\multicolumn{2}{@{}l}{\emph{Functional test programs}} \\
\addlinespace[2pt]
Conditional swap & Swap two values according to a secret comparison \\
Search and counting & Bounded search; bounded threshold count \\
Secret-indexed access & Secret-indexed read; secret-indexed update \\
Branches and memory & Lookup using a branch-selected index; update with a value selected by nested branches \\
Loops and memory & Secret-indexed lookup and accumulation within a loop \\
\bottomrule
\end{tabularx}
\par\vspace{2pt}
\begin{minipage}{\columnwidth}
\footnotesize\raggedright
The first three rows define $3\times3\times3-1=26$ mechanism configurations, excluding the all-none case. Together with 8 functional programs, the suite contains 34 programs. Generated loop variants have secret-dependent termination. Generated programs use arithmetic, bitwise, or comparison expressions and input arrays with a public length of 4 or 8 words per party.
\end{minipage}
\end{table}

For each test input, the original and transformed RV32 programs running on Spike produce the same complete outputs as an independent reference interpreter. Representative programs also execute in the Alice/Bob runtime, where their decoded outputs match the reference results. We additionally introduce missing or invalid loop bounds and scan extents, unsupported control or addressing, and malformed commit structures into otherwise valid programs; all such cases are rejected before execution.

For every RV2GC benchmark in the performance evaluation, we also compare the complete secure-execution output with a Spike reference execution using the same inputs; all outputs agree. The next section further checks whether control flow and memory addresses in the final binaries are independent of private inputs.

\subsection{Control-Flow and Memory-Address Verification}
\label{sec:binsec-method}

To check that compiled programs satisfy the execution requirement in Section~\ref{sec:threat-model}, we use BINSEC 0.11.3 to analyze whether control flow and memory addresses in the final binaries depend on private inputs~\cite{DanielBR23Binsec}. For a fixed program and public configuration, private inputs may affect the computed results, but must not change the publicly observable instruction and address sequences.

We extend BINSEC's instruction semantics to model RV2GC's custom conditional-execution instructions, \rvinst{pushbr}, \rvinst{updatebr}, and \rvinst{popbr}, together with protected state updates. Conditional stores retain their publicly observable address accesses; the protected predicate controls only the committed value.

For each configuration, we initialize the declared public state and make both parties' full private-input regions symbolic, then use relational symbolic execution to check trace independence. Analysis covers execution from entry to \texttt{main} through its return. It also checks that predicate values are Boolean and that conditional-stack operations are balanced and within capacity. Verification succeeds only when the complete analysis establishes trace independence, with no unresolved checks or failed assertions.

The extended instruction model agrees with independent symbolic output specifications and concrete Spike execution results with Boolean predicates. We also check the analysis using control programs with secret-dependent branches and secret-indexed accesses: untransformed programs expose the corresponding control-flow or address dependencies, whereas transformed programs pass verification. Invalid predicate values, stack operations, and output specifications trigger the corresponding assertion failures.

All 37 benchmark configurations used in our evaluation pass final-binary analysis. Under the declared public configurations, execution boundaries, and extended instruction model, their control flow and memory addresses are independent of private inputs. These results provide binary-level evidence for the public-execution property required by the preceding privacy argument.

\section{Evaluation}

We evaluate the three design choices through circuit size and measured runtime, stateful composition over a fixed library, and the runtime costs of protected execution. Comparisons with other programmable systems and function-specific circuits then examine how different approaches trade reuse against execution costs.

\subsection{Experimental Setup}
\label{Experimental Setup}

Experiments run on an Intel Core i7-14650HX with 15~GiB of memory and Ubuntu~22.04, with TCP loopback communication between the parties.\footnote{The main ARM2GC comparison and scaling study use one measurement per configuration, from immediately after Bob's launch until both processes exit. Four-system workload and function-specific comparisons report three-run medians for time and communication. The former uses process elapsed time for RV2GC/ARM2GC, the longer party time for Wang, and evaluator time for VISA; the latter uses a synchronized interval after program loading, covering input encoding, OT, evaluation, and output.} RV32 programs are compiled with RISC-V GCC~15.2.0 at \texttt{-O2}, and instruction circuits are synthesized with Yosys~0.9. RV2GC and ARM2GC share the underlying garbling implementation and use 128-bit wire labels.

We use matched inputs and check complete outputs. We report online execution time and the total application-level bytes sent by both parties. Online time excludes compilation, schedule generation, and circuit synthesis.

\subsection{Circuit Boundary and Measured Runtime}
\label{sec:overall-performance}

We first compare reusable circuit topologies, then evaluate runtime using the fixed 4,096-byte Amber configuration.

\subsubsection{Static Boundary Size}
\label{sec:explaining-gains}

Table~\ref{tab:static-boundary} compares the gate counts of ten RV2GC protected-circuit topologies with the 256-byte and 4,096-byte ARM2GC Amber processors. The multiplication circuit is the largest of the ten listed RV2GC circuits, with 3,386 gates, corresponding to 2.044\% and 0.393\% of the gate counts of the two Amber processors, respectively.

\begin{table}[!tb]
  \scriptsize
  \centering
  \setlength{\tabcolsep}{1.5pt}
  \renewcommand{\arraystretch}{0.94}
  \caption{RV2GC protected-circuit sizes relative to complete ARM2GC Amber processors.}
  \label{tab:static-boundary}
  \begin{tabular*}{\columnwidth}{@{\extracolsep{\fill}}lllrrr@{}}
    \toprule
    \multirow[c]{2}{*}{System}
      & \multirow[c]{2}{*}{Class}
      & \multirow[c]{2}{*}{Circuit}
      & \multirow[c]{2}{*}{Gates}
      & \multicolumn{2}{c}{Relative to ARM2GC} \\
    \cmidrule(lr){5-6}
      & & & & \multicolumn{1}{c}{Amber 256-B} & \multicolumn{1}{c}{Amber 4096-B} \\
    \midrule
    \multirow{10}{*}{RV2GC}
      & \multirow{3}{*}{Arithmetic} & add & 244 & 0.147\% & 0.028\% \\
      &                              & sub & 244 & 0.147\% & 0.028\% \\
      &                              & mul & 3,386 & 2.044\% & 0.393\% \\
    \cmidrule(lr){2-6}
      & \multirow{2}{*}{Logical} & and & 32 & 0.019\% & 0.004\% \\
      &                           & or  & 32 & 0.019\% & 0.004\% \\
    \cmidrule(lr){2-6}
      & \multirow{2}{*}{Shift} & sll & 419 & 0.253\% & 0.049\% \\
      &                         & sra & 444 & 0.268\% & 0.052\% \\
    \cmidrule(lr){2-6}
      & \multirow{3}{*}{Compare} & slt & 185 & 0.112\% & 0.021\% \\
      &                            & sltu & 183 & 0.110\% & 0.021\% \\
      &                            & swapgt & 256 & 0.154\% & 0.030\% \\
    \midrule
    \multirow{2}{*}{ARM2GC} & \multicolumn{2}{c}{Amber 256-B} & 165,696 & 100\% & --- \\
                              & \multicolumn{2}{c}{Amber 4096-B} & 860,536 & --- & 100\% \\
    \bottomrule
  \end{tabular*}
\end{table}

\subsubsection{Runtime Comparison with ARM2GC}

\begin{figure}[!tb]
  \centering
  \includegraphics[width=\columnwidth]{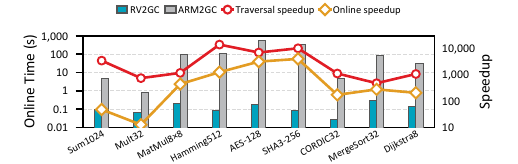}
  \vspace{-2pt}
  \caption{Comparison of RV2GC and ARM2GC across nine matched workloads. Bars show online execution time; lines show traversal and online execution speedups.}
  \label{fig:boundary-impact}
\end{figure}

Table~\ref{tab:boundary-impact} compares ARM2GC and RV2GC across nine benchmarks. ARM2GC uses its 4,096-byte Amber processor with SkipGate, and RV2GC uses its fixed instruction-circuit library. The two secret-dependent workloads use protected commits and fixed scans in RV2GC, and ordinary branches and indexed accesses in ARM2GC. Figure~\ref{fig:boundary-impact} presents the corresponding runtime and traversal speedups.\footnote{Gate-traversal time reports the remaining evaluator time after subtracting protected-gate-handler time. The RV2GC measurement also includes garbled-table reception.}

\begin{table}[!tb]
  \scriptsize
  \centering
  \setlength{\tabcolsep}{1.15pt}
  \renewcommand{\arraystretch}{0.94}
  \caption{Comparison of RV2GC and ARM2GC across nine matched workloads in terms of gate traversal time and complete online execution time. Speedups are defined as $T_{\mathrm{ARM2GC}}/T_{\mathrm{RV2GC}}$ and computed from unrounded times.}
  \label{tab:boundary-impact}
  \begin{tabular*}{\columnwidth}{@{\extracolsep{\fill}}lrrrrrr@{}}
    \toprule
    \multirow[c]{2}{*}{Workload}
      & \multicolumn{2}{c}{Gate Traversal (s)}
      & \multirow[c]{2}{*}{Speedup}
      & \multicolumn{2}{c}{Online Execution (s)}
      & \multirow[c]{2}{*}{Speedup} \\
    \cmidrule(lr){2-3} \cmidrule(lr){5-6}
      & RV2GC & ARM2GC & & RV2GC & ARM2GC & \\
    \midrule
    \multicolumn{7}{l}{\itshape Public control flow} \\
    Sum1024          & 0.001 & 3.832   & 3,474.94$\times$  & 0.095 & 4.611   & 48.40$\times$ \\
    Mult32           & 0.001 & 0.591   & 751.59$\times$    & 0.065 & 0.829   & 12.81$\times$ \\
    MatMul8$\times$8 & 0.065 & 77.125  & 1,188.12$\times$  & 0.218 & 96.958  & 445.30$\times$ \\
    Hamming512       & 0.007 & 103.917 & 14,138.28$\times$ & 0.088 & 111.526 & 1,265.92$\times$ \\
    AES-128          & 0.075 & 528.478 & 7,067.54$\times$  & 0.182 & 578.264 & 3,185.91$\times$ \\
    SHA3-256         & 0.032 & 326.066 & 10,219.83$\times$ & 0.091 & 368.845 & 4,042.58$\times$ \\
    CORDIC32         & 0.004 & 4.563   & 1,127.02$\times$  & 0.029 & 5.126   & 174.32$\times$ \\
    \midrule
    \multicolumn{7}{l}{\itshape Secret-dependent control flow} \\
    MergeSort32      & 0.143 & 68.740 & 482.10$\times$   & 0.311 & 87.720 & 281.69$\times$ \\
    Dijkstra8        & 0.026 & 27.712 & 1,080.71$\times$ & 0.149 & 31.047 & 207.99$\times$ \\
    \bottomrule
  \end{tabular*}
\end{table}

The measured online runtime speedups range from 12.81$\times$ to 4,043$\times$. Substantial speedups also occur in the seven public-control benchmarks, with SHA3-256 giving the largest ratio. These gains therefore extend beyond workloads requiring secret-control-flow transformations.

ARM2GC also exploits public control through ARM conditional execution~\cite{DBLP:conf/dac/SonghoriRHSK19}. RV2GC's structural distinction is that a public operation directly selects its execution path without processor-wide gate classification.

\begin{figure*}[!b]
  \centering
  \setlength{\subfigcapskip}{-1pt}
  \setlength{\subfigbottomskip}{0pt}
  \includegraphics[width=0.57\textwidth]{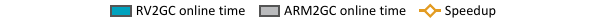}\\[-1.0ex]
  \subfigure[Secret-Dependent Branch]{\label{fig:protected-scaling-branch}\includegraphics[width=0.325\textwidth]{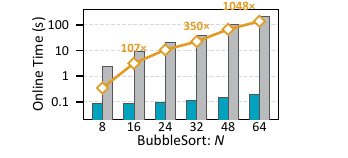}}\hfill
  \subfigure[Secret-Dependent Loop]{\label{fig:protected-scaling-loop}\includegraphics[width=0.325\textwidth]{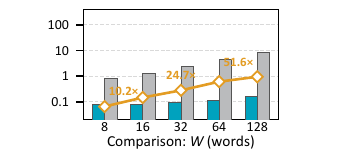}}\hfill
  \subfigure[Secret-Indexed Memory]{\label{fig:protected-scaling-memory}\includegraphics[width=0.325\textwidth]{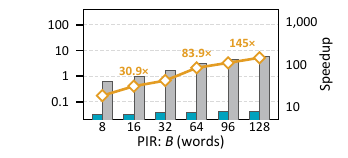}}
\caption{Runtime scaling of BubbleSort, Comparison, and PIR in RV2GC and ARM2GC. Bars show online execution time; lines show ARM2GC-to-RV2GC speedups. The scale variables are public array length, loop bound, and scan size, respectively.}
  \label{fig:protected-scaling}
\end{figure*}

\subsection{Stateful Execution with a Fixed Circuit Library}
\label{sec:compiled-scalability}

We examine two properties: whether different applications reuse the same fixed instruction-circuit library, and whether separately invoked circuits continuously pass and update architectural state within long-running programs with different structures.

\subsubsection{Fixed Library and Evaluated ISA Scope}

All RV2GC programs in this evaluation use the same fixed instruction-circuit library. The dispatcher accepts 34 RV32 opcodes across protected-circuit, public-computation, and label-only paths. The evaluated corpus dynamically exercises 31 of them and contains no opcode outside the accepted set. The evaluation covers RV2GC's implemented RV32 integer subset; floating-point operations, atomics, and exception semantics are outside the evaluated scope.

\subsubsection{Long-Running Stateful Programs}

We evaluate two programs listed in Table~\ref{tab:program-scale-instances}: Mini-Transformer with attention, residual connections, and a feed-forward network, and Mini-ResNet with convolution, residual blocks, pooling, and a classifier. Both programs have public model structures and loop bounds. Their inputs and weights use signed int8 values, with int32 accumulation.

\begin{table}[!tb]
  \scriptsize
  \centering
  \setlength{\tabcolsep}{1.5pt}
  \caption{Long-running integer programs executed with the same fixed instruction-circuit library.}
  \label{tab:program-scale-instances}
  \begin{tabularx}{\columnwidth}{@{}
    >{\hsize=0.96\hsize\linewidth=\hsize\raggedright\arraybackslash}X
    >{\hsize=1.68\hsize\linewidth=\hsize\raggedright\arraybackslash}X
    >{\hsize=0.84\hsize\linewidth=\hsize\raggedleft\arraybackslash}X
    >{\hsize=0.52\hsize\linewidth=\hsize\raggedleft\arraybackslash}X@{}}
    \toprule
    Program & Computation Structure & Dynamic Inst. & Online (s) \\
    \midrule
    Mini-Transformer & Attention, residual, FFN & 3,411,714 & 80.396 \\
    Mini-ResNet & Convolution, residual, pooling & 11,022,541 & 178.015 \\
    \bottomrule
  \end{tabularx}
\end{table}

Mini-Transformer and Mini-ResNet execute 3.41 and 11.02 million scheduled instructions, respectively, using the same fixed library. All ten int32 output logits of each program match the Spike reference results. Their public schedules occupy 87.76 MiB and 283.35 MiB, respectively. These sizes show the storage cost of representing a long execution as a public schedule. Both programs also pass binary trace verification at their full public sizes (Section~\ref{sec:binsec-method}).

\subsection{Runtime Scaling of Secret-Dependent Execution}
\label{sec:protected-scaling}

Secret-dependent branches, loop termination, and memory indices require computation under a public execution sequence. This experiment examines how RV2GC's runtime changes as such work grows and whether its performance advantage over ARM2GC persists. We increase the problem sizes of three benchmarks to study conditional swaps, secret-dependent loops, and secret-indexed reads.

For BubbleSort, we increase the array length from 8 to 64, adding conditional swaps whose outcomes depend on protected comparisons. For Comparison, we increase the comparison length from 8 to 128; private data determines the logical stopping point, while the public schedule continues over the prescribed range. For PIR, every read scans 8--128 32-bit entries and uses protected selection to return the secret-indexed entry. These workloads exercise the mechanisms in Section~\ref{sec:secret-dependent-execution}. Figure~\ref{fig:protected-scaling} reports their runtimes and speedups.

All points in the secret-dependent execution scaling study use the same Amber configuration: 4,096-byte code memory, 128-word input memories for both parties, 64-word output memory, and 64-word stack memory.

Over the evaluated sizes, ARM2GC's runtime grows faster, and RV2GC's speedup increases with size in all three benchmarks, reaching 1,048$\times$, 51.6$\times$, and 145$\times$ at their largest configurations. Thus, as conditional swaps, loop iterations, and scan work increase, RV2GC's combination of public scheduling and protected state updates retains a runtime advantage over ARM2GC.

\begin{figure*}[!t]
  \centering
  \setlength{\subfigcapskip}{-1pt}
  \setlength{\subfigbottomskip}{0pt}
  \includegraphics[width=0.98\textwidth]{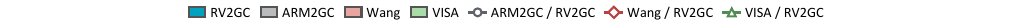}\\[-1.0ex]
  \subfigure[Online Time]{\label{fig:four-system-online}\includegraphics[width=0.325\textwidth]{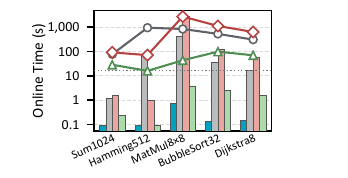}}\hfill
  \subfigure[Communication]{\label{fig:four-system-communication}\includegraphics[width=0.325\textwidth]{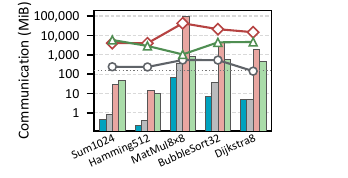}}\hfill
  \subfigure[Program Data]{\label{fig:four-system-program-data}\includegraphics[width=0.325\textwidth]{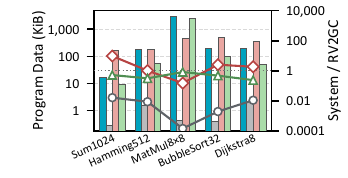}}
  \caption{Comparison of RV2GC, ARM2GC, Wang, and VISA across five workloads. The three panels report online execution time, communication, and Program Data, with bars for absolute values and lines for system-to-RV2GC ratios.}
  \label{fig:four-system-comparison}
\end{figure*}

\begin{figure*}[!b]
  \centering
  \setlength{\subfigcapskip}{-1pt}
  \setlength{\subfigbottomskip}{0pt}
  \includegraphics[width=0.98\textwidth]{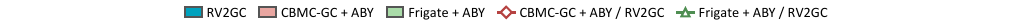}\\[-1.0ex]
  \subfigure[Online Time]{\label{fig:function-specific-online}\includegraphics[width=0.325\textwidth]{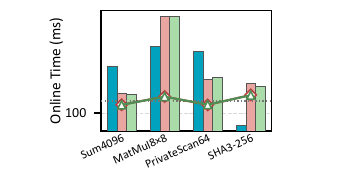}}\hfill
  \subfigure[Communication]{\label{fig:function-specific-communication}\includegraphics[width=0.325\textwidth]{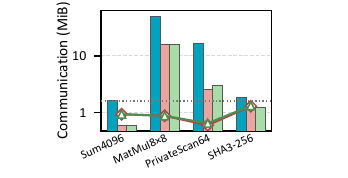}}\hfill
  \subfigure[Circuit or Schedule Size]{\label{fig:function-specific-size}\includegraphics[width=0.325\textwidth]{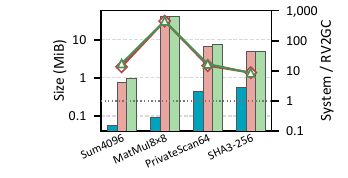}}
  \caption{Comparison of RV2GC with function-specific circuits generated by CBMC-GC and Frigate across four computations. The three panels report online execution time, communication, and circuit-or-schedule size, with bars for absolute values and lines for CBMC-GC+ABY/RV2GC and Frigate+ABY/RV2GC ratios.}
  \label{fig:function-specific-comparison}
\end{figure*}

\subsection{Tradeoffs with Other Programmable GC Systems}
\label{sec:cross-system-comparison}

We compare five computations across the four systems and sweep the execution length of a common stateful test program to identify resource growth. The comparisons cover runtime, communication, program artifacts, and derived state.

\subsubsection{Workload-Level Comparison}

Figure~\ref{fig:four-system-comparison} compares five computations across RV2GC, ARM2GC, Wang, and VISA.

For each workload, ARM2GC uses the smallest evaluated Amber configuration that accommodates the program.\footnote{Sum1024, MatMul8$\times$8, and BubbleSort32 use the 256-byte Amber; Hamming512 and Dijkstra8 use the 2,048-byte Amber. The main comparison in Table~\ref{tab:boundary-impact} uses a fixed 4,096-byte configuration.} MatMul8$\times$8 and BubbleSort32 use fixed-work versions shared by all four systems, distinct from those in the main ARM2GC and function-specific comparisons. Wang uses its supported MIPS-I subset with a public step bound; VISA uses assembly programs and native fragment construction.\footnote{We add complete-output extraction to Wang and VISA and correct Wang's store-source register-liveness mask.}

\emph{Program Data} denotes the program-specific representation and initialization or configuration data retained after preparation for later executions, excluding the fixed substrate shared across programs.

RV2GC achieves runtime speedups of 12.6--731$\times$ over ARM2GC and 11.0--4,021$\times$ over Wang across the five workloads. RV2GC is 2.52--18.7$\times$ faster than VISA on four workloads and has comparable runtime on Hamming512. RV2GC communicates less than Wang and VISA on all five workloads, while ARM2GC communicates less on Dijkstra8.

Program Data captures each architecture's program-specific representation. Because its public schedule is program specific, RV2GC retains more than ARM2GC and VISA on all five workloads, less than Wang on four, and more than Wang on MatMul8$\times$8. ARM2GC's complete processor circuit is fixed across programs and therefore excluded from this metric.

\subsubsection{Stateful Execution-Length Scaling}

StatefulChain32 repeatedly updates two private 32-bit words within the same public loop. We vary the iteration count $K\in\{64,256,1024,4096,16384\}$ to isolate the effect of execution length. Each iteration performs two additions and one XOR modulo $2^{32}$. Every port retains a backward loop in its binary or assembly representation rather than replicating the loop body $K$ times. For RV2GC, preparation produces a public instruction schedule containing the operations from all $K$ iterations. At each point, we record structural quantities, program-object sizes, and communication from one two-party execution.

Table~\ref{tab:stateful-chain-scaling} summarizes how resource usage changes with execution length: RV2GC grows a $4K+12$-entry public schedule while peak derived state remains 1,383 bytes; ARM2GC grows processor cycles; Wang keeps generated CPU and memory structures fixed while communication grows approximately linearly; and VISA keeps its program and control-flow graph fixed while active entries and derived state grow.

\begin{table}[!tb]
  \scriptsize
  \centering
  \setlength{\tabcolsep}{1.5pt}
  \caption{Comparison of RV2GC, ARM2GC, Wang, and VISA on StatefulChain32 in terms of architecture-specific resource growth with execution length.}
  \label{tab:stateful-chain-scaling}
  \begin{tabular*}{\columnwidth}{@{\extracolsep{\fill}}llll@{}}
    \toprule
    System & Grows with $K$ & Scaling Law & Derived State (Bytes) \\
    \midrule
    RV2GC & Public schedule & $107K+337$ bytes & 1,383 (constant) \\
    ARM2GC & Processor cycles & $7K+39$ cycles & 233 (constant) \\
    Wang & Communication & $(0.241K+2.23)$ MiB & 19,536 (constant) \\
    VISA & Active entries & $4K-1$ entries & $40K+2{,}201$ \\
    \bottomrule
  \end{tabular*}
  \par\vspace{2pt}
  \begin{minipage}{\columnwidth}
    \scriptsize\raggedright
    \emph{Derived State} reports the size of program-derived data structures tracked by each implementation, such as schedule-parsing buffers or active-entry lists; it does not represent total process memory.
  \end{minipage}
\end{table}

\subsection{Tradeoffs with Function-Specific Circuits}
\label{sec:function-specific-comparison}

Whole-function specialization generates a complete secure circuit for each computation, whereas RV2GC reuses a fixed instruction-circuit library across programs. The comparison covers aggregation, matrix multiplication, secret-indexed lookup, and hashing. We execute CBMC-GC- and Frigate-generated circuits under the same ABY/Yao backend~\cite{DemmlerSZ15} and compare them with RV2GC.

Circuit or schedule size counts the program-specific complete circuit or public schedule, excluding shared structures.

As shown in Figure~\ref{fig:function-specific-comparison}, RV2GC has lower online execution time than both function-specific implementations on MatMul8$\times$8 and SHA3-256, but higher time on Sum4096 and PrivateScan64. Both function-specific implementations communicate less in all four cases. For each workload, the smaller of the two function-specific circuit files is 8.55--449.72 times the size of the corresponding RV2GC public schedule.

Fixed topology reuse removes per-program instruction-circuit synthesis; program preparation generates and stores a public schedule.

\section{Related Work}
\label{sec:related work}

\noindent\textbf{Garbled processors and reusable execution structures.}
TinyGarble develops compact and scalable sequential garbled circuits, GarbledCPU realizes a MIPS garbled processor, and ARM2GC applies SkipGate~\cite{DBLP:conf/sp/SonghoriHS0K15,songhori2016garbledcpu,DBLP:conf/dac/SonghoriRHSK19}. RV2GC moves the reusable boundary from the complete processor topology used by GarbledCPU and ARM2GC to a fixed library of RISC-V instruction circuits, keeping fetch and decode outside the repeatedly traversed boundary. ARM2GC is therefore its closest architectural baseline.

\noindent\textbf{Secure machine code and variable ISAs.}
Wang et al. execute MIPS binaries with a secret program counter and oblivious instruction and data access. Their implementation uses statically generated step-indexed instruction banks and operation-set-specialized CPU circuits~\cite{DBLP:conf/esorics/WangGMK16}. VISA derives per-program fragment circuits and active-set-specific stacks, using Stacked Garbling for protected fragment selection and EpiGRAM for memory access~\cite{DBLP:conf/ccs/YangPHK23}. RV2GC uses a fixed RISC-V instruction-circuit library with public, secret-invariant instruction and memory-access sequences. Program changes affect its public schedule and initialization data rather than circuit topology.

\noindent\textbf{Function-specific GC compilation.}
Secure-computation toolchains such as Fairplay, PCF, CBMC-GC, Frigate, and Obliv-C use languages, intermediate representations, or compiler paths to specialize secure computation to a target program or function~\cite{DBLP:conf/uss/MalkhiNPS04,DBLP:conf/uss/KreuterSMB13,DBLP:conf/cc/FranzHKSV14,DBLP:conf/eurosp/MoodGCBT16,ZahurEvansOblivC15}. Our evaluation compares RV2GC's fixed ISA-level library with function-specific circuits from CBMC-GC and Frigate.

\noindent\textbf{Complementary memory and GC techniques.}
ObliVM integrates oblivious data structures and Garbled RAM targets sublinear protected memory access, whereas RV2GC uses a fixed scan within a public block~\cite{DBLP:conf/sp/LiuWNHS15,DBLP:conf/eurocrypt/GentryHLO0W14}. Hardware/software co-design, logic synthesis, specialized lookup constructions, and arithmetic garbling provide techniques complementary to RV2GC's boundary reconstruction~\cite{DBLP:conf/isca/MoGR23,DBLP:journals/tifs/PohleAP24,DBLP:journals/tches/SanderBBE25,DBLP:conf/iccad/YuM23,DBLP:conf/eurocrypt/HeathKN24}. Within its operation-level boundary, RV2GC applies instruction-specific garbling-free optimizations where eligible.

\section{Conclusion}

RV2GC reconstructs the garbling boundary around reusable RISC-V instruction circuits, connects independently invoked instruction circuits and garbling-free paths through persistent label state, and uses protected commits to let secret-dependent choices govern state updates. These mechanisms enable stateful execution without repeated complete-processor traversal. RV2GC is faster than ARM2GC across the evaluated benchmarks. Its public schedules occupy program-specific storage, secret-indexed accesses require fixed scans, and communication exceeds that of the evaluated function-specific circuits.

\IEEEtriggeratref{1}
\IEEEtriggercmd{%
  \setlength{\baselineskip}{0.98\baselineskip}%
  \IEEEtriggeratref{17}%
  \IEEEtriggercmd{\balance}}
\bibliographystyle{IEEEtranS}
\bibliography{refs}
\end{document}